\documentclass[acmsmall,authorversion,nonacm]{acmart}   

\AtBeginDocument{%
  \providecommand\BibTeX{{%
    \normalfont B\kern-0.5em{\scshape i\kern-0.25em b}\kern-0.8em\TeX}}}

\usepackage{url}
\usepackage{doi}

\usepackage{currfile}
\usepackage{textcomp}

\usepackage{enumitem}

\begin{document}

\title{Towards Blockchain-enabled Open Architectures for Scalable Digital Asset Platforms}

\author{Denis Avrilionis}
\affiliation{%
\institution{Compellio SA}
\city{Luxembourg}
\country{Luxembourg. }
\href{mailto:denis@compell.io}{denis@compell.io}
}

\author{Thomas Hardjono}
\affiliation{%
\institution{MIT Connection Science \& Engineering}
\city{Cambridge, MA}
\country{USA. }
\href{mailto:hardjono@mit.edu}{hardjono@mit.edu}
}

\begin{abstract}
Today there is considerable interest in deploying blockchains and decentralized ledger technology as a means to address the deficiencies of current financial and digital asset infrastructures.
The focal point of attention in many projects on digital asset and cryptocurrency is centered around blockchain systems and smart contracts. Many projects seek to make the blockchain as the centerpiece of the new decentralized world of finance. However, several roadblocks and challenges currently face this predominant blockchain-centric view.
In this paper we argue that the proper and correct perspective on decentralized economy should be one that is asset-centric, where the goal should be the consistent lifecycle management of assets in the real-world with their digital representation on the blockchain.
We introduce the notion of the {\em digital twin} to capture 
the relationship between a real-world asset and its on-chain representation. 
A {\em digital twin container} is utilized to permit off-chain state persistence and 
on-chain state traceability, where the container can be deployed on the blockchain as well as on traditional application servers. 
The digital twin container becomes the bridge between legacy infrastructures and 
the newly emergent blockchain infrastructures, 
permitting legacy systems to interoperate consistently with blockchain systems.
We believe this asset-centric view to be the correct evolutionary direction 
for the nascent field of blockchains and decentralized ledger technology.\\
~~\\
{\bf Date: \today}

~~
\end{abstract}





\maketitle


\section{Introduction}

The emergence of
blockchains and smart contracts technology to enable the tokenization of financial assets has propelled great interest in smart contracts as a potential paradigm for decentralized finance. Like the various financial networks that today exist in different trading blocks, we believe that multiple blockchain networks will emerge in the future and that different forms of tokenized assets will also be deployed. This raises several challenges with regards to the proper place of blockchain and distributed ledger technology (DLT) within the landscape of financial computing generally and within open architecture more specifically. A key challenge in achieving open and accessible platforms as part of the vision of decentralized finance is that of the integration of blockchain and DLTs with legacy systems and infrastructures.

For digital assets to freely flow among such heterogeneous systems 
we must ensure that the overall computation paradigm for 
a conceptual ``network of networks'' offers openness and trust. 
In particular, the ability to maintain the consistency of the state of assets 
is a key capability requirement of all systems participating in such a globally interconnected world.

The current design of blockchain and smart contracts indirectly leads to the creation of silos, both data and assets. This is because, among others, the current smart contracts paradigm has a very constrained view of the world and requires particular components (known as "oracles") to interact with other external systems. These oracles have to "pull in" data from the external world and make it available on the ledger of the blockchain. 

We believe that a new and broader computational paradigm is needed to underpin and encompass blockchain-based digital asset transactions. This new paradigm must permit assets to flow into and out of blockchains and legacy systems seamlessly. The new paradigm must also allow off-chain data and other asset-related state information in digital form to be reachable by the smart contract. Moreover, computations occurring in legacy systems must also be possible in coordination with smart contracts.  

This implies that good design principles are needed for the following: 
(a) to minimize on-chain state information, 
(b) to connect/link on-chain state to richer off-chain information (maintained by an asset state "custodian" entity), and 
(c) to employ an active function/entity that coordinates the exchange of assets among heterogeneous systems.

When an application program (e.g., an asset trading application) 
seeks to modify the state of an asset, it needs to ensure that the real-world asset (e.g., real estate ownership certificate) 
and the digital representation of the asset on the blockchain (e.g. materialised by smart contract code) are synchronized. To address this challenge and central to our approach, is the ability of a system 
-- either blockchain-based or non-blockchain based -- 
to {\em mirror} the state of real-world assets (both digital or physical assets) with their digital ``twin'' representations. 
This continuous synchronization is performed to prevent changes in the asset ownership on the blockchain from being conducted without a corresponding state-change in the real world asset. 

To ensure the correct and consistent state on both worlds, 
we introduce the notion of the asset {\em digital twin}
and a corresponding mediating software layer -- which we refer to
as the digital twin container (see Section~\ref{Sec:AssetCentric}) --
sits in between the two worlds (on-chain and off-chain).

To achieve this new paradigm, 
we also need to enhance current architectures to include functions 
related to:
(i) the management of the asset repository (off-chain) and 
(ii) related to the verification of the consistency
of on-chain state information (which we refer to as {\em state verification} henceforth). 

Related to maintaining the consistency between on-chain/off-chain states,
we borrow the classic database notion of the {\em Logical Unit of Work} (LUW)
to mean the set of sub-transactions that must complete
across various other computers in order to complete the work.
The LUW corresponds to the traditional attribute of a transaction
in a distributed database system that must satisfy the ACID properties
(Atomicity, Consistency, Isolation, Durability).
The LUW is a way to unify the mixed world of on-chain and off-chain state management,
where a single top-level transaction may in fact be composed of a spread of
sub-transactions~\cite{HardjonoLipton-IEEETEMS-2019}
that may span across multiple computers systems.
Here, each sub-transaction can be either off-chain or on-chain,
and each must reach final a commitment stage
in order for the parent top-level transaction to be considered
as committed or settled.

In order to combine off-chain persistence of state information
and the on-chain timestamping effect of the blockchain upon the digital twin
(i.e., the on-chain digital representation of the asset),
we introduce the approach of the {\em Digital Twin Container} (DT-Container).
A given DT-Container can be deployed on a blockchain or 
within a traditional application server,
and it can host computational  units either as smart contracts in the former case, 
or as software packages in the latter case. 
By following the classic and proven 
{\em Design by Contract} approach~\cite{Meyer1994} 
the DT-Container checks for pre-/post-conditions and invariants
before and after processing the state of digital twins. 
Invalid properties may lead to non-execution of the code
or to the invalidation of the computation results.
This is in order always to maintain consistency of the state of digital twins
with respect to the corresponding assets in the real world.

By using this approach, 
the computation that modifies the state of assets 
can be performed not only by smart contracts but also 
by other applications running in non-blockchain systems 
(including legacy systems). 
Besides the usual \texttt{construct / read / update / destruct} operations, 
the DT-Container is also responsible for negotiating and executing 
the transfer of assets to and from other external systems 
via a fully transactional digital twin, 
thus enabling free-flow of assets in a consistent manner 
in the context of global LUWs.

With regards to digital assets,
today there is tremendous interest in digitizing assets and 
making these assets accessible and tradeable via decentralized systems, 
such as blockchains and DLT networks. 
Adding to this mix -- and often introducing confusion -- is the fact that 
some blockchain systems possess the technical capability to 
generate and manage endogenous tokens 
(e.g., ERC-20 compliant tokens on Ethereum~\cite{VogelstellerButerin2015}) 
that may be considered a new class of assets. 

In the current work we seek to bridge the world of off-chain assets 
with technical capabilities that are available on-chain. 
As such, we distinguish among the following types of assets:
\begin{enumerate}[label=(\alph*), topsep=8pt,itemsep=1pt, partopsep=4pt, parsep=4pt]

\item Legally recognized assets in non-digital representation (off-chain): 
These are assets that exist off-chain, usually represented by 
depository receipts of certificates.

\item Digitized equivalents of legally recognized assets (off-chain): 
These are the local digitized versions of (a) above where, 
for example, a depository receipt (paper) has been digitally 
represented in the local database (or local ledger) of the depository institution 
    (e.g., see~\cite{DTCC-Project-Whitney-2020,DTCC-Project-ION-2020}).

\item On-chain tokenized assets:  This includes exogenous digitized assets 
(e.g. (b) above) that have been tokenized, as well as ledger-endogenous tokens.

\end{enumerate}

In our approach, Digital Twins are digital representations of assets, 
be they off-chain digitized assets (case of (b) above) or, native on-chain assets (case of (c) above).
    
The topic of digital assets, tokens, and their definition are out-of-scope for the current work,
and has been treated elsewhere 
(e.g. see~\cite{LiptonTreccani2021-book,Pentland2021Barrons} for 
a discussion on the economic impact of tokenized assets, see~\cite{EUBOF2020-NFT} regarding NFTs
and see~\cite{ECB2019-CryptoAssets,GDF2019-Taxonomy,ITSA2021-TokenClassification} 
for industry efforts
to standardize the definitions of tokens and related assets).

\section{Limitations of the Current Smart Contract Paradigm}

An important core idea underlying the notion of smart contracts today
is the sharing of common functions and state information (data) on the ledger. 
Both of these computing constructs are not new inventions 
but rather originate from the field of object-oriented programming 
(e.g., Smalltalk, C++, Java). 
Indeed, the work of~\cite{Herlihy2019-CACM} immediately refers
these constructs 
as objects in the classical object-oriented programming sense, 
where the function is typically referred to a methods 
and the state as variables in the object. 

The innovation introduced by smart contracts as exemplified by
the Ethereum platform~\cite{Buterin2014} 
is the replication of the function/state modules 
across all nodes of the blockchain platform and 
the tight coupling with the ledger as the mechanism 
to store the output of the smart contract~\cite{Solidity2014}. 
This means that the smart contract code and 
its resulting output are readable by any other 
smart contract on the same platform. 

\begin{figure}[t]
\centering
\includegraphics[width=0.7\textwidth, trim={0.0cm 0.0cm 0.0cm 0.0cm}, clip]{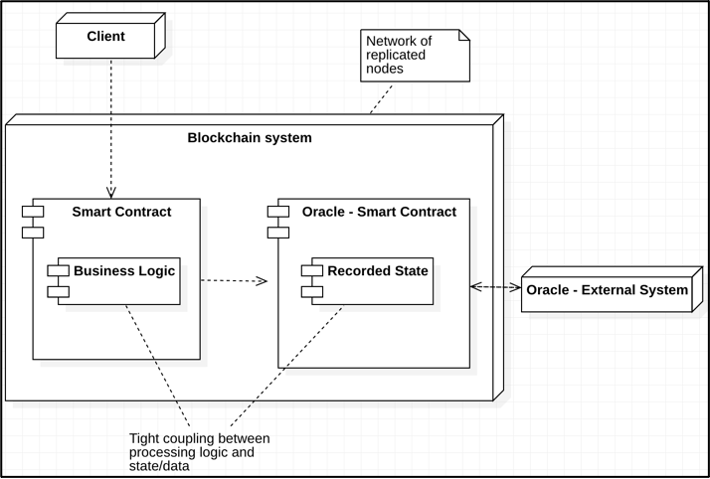}
\caption{Illustration of the tight coupling between smart contracts and state/data recorded on the replicated nodes of the blockchain}
\label{fig:figure1label}
\end{figure}

The author of a smart contract can design the logic of the contract object 
to be such that when a piece state-data (such as a digital asset representation) 
is being used (accessed) by one node
(e.g., instance X in node Y at location Z)
that the node has exclusive use of it,
by virtue of the consensus mechanism being utilized by the nodes
(e.g., proof of stake).
Once the object instance at the node
terminates execution and produces changes to the state-data, 
that new state-data is appended to the ledger
(i.e., new block propagated to all nodes). 
Given the same input data, any copy of the smart contract (i.e., on any node)
must produce the same output.

The smart contract code today has a limited world-view
because unlike other types of software
it cannot rely on information fetched from sources outside the ledger --
which may be ever-changing (ephemeral) or become unavailable very soon after.  
This means that some time $t$ later
other peer nodes may not have access
to the identical data from the external source, 
and therefore will be unable to validate the veracity of the information.

Due to the above limitations, 
special components -- referred to as Oracles -- are needed to make these externally-sourced data available on the ledger 
by writing these data onto the blocks of the ledger.\footnote{It is worth
noting that Oracles themselves can implement 
simple smart contracts to make external data available 
to other ``computational'' (non-oracle) smart contracts 
in the same blockchain system.}
However, this implies that an increasing amount of data 
must be added to the ledger of the blockchain so that smart contracts 
can read this data (see Figure 1). 
Over time, this leads to the accumulation of data on the ledger,
potentially leading the ledger becoming a silo of data over time.

\section{The Oracle Problem Revisited}

Blockchains have been designed with the fundamental assumption 
that the state of the overall system is defined by the combination of:
(a) the content of wallets 
(i.e. digital tokens bound to public keys),
and (b) the state of smart contracts interpreted by the system. 
This closed view often leads to the distinction 
between on-chain state (the state of the blockchain ledger system) 
and off-chain state (the state of the outside world). 
Besides technical aspects related to synchronizing non-blockchain systems 
with on-chain digital assets state,
several issues are hindering the road to
blockchain-based digitization/tokenization~\cite{Uzsoki2019}. 
These include the absence of the legal framework and regulations 
to confer legal value to on-chain records,
the lack of legal accountability for those records in the context of
data privacy~\cite{GDPR} and corporate confidentiality protection.
In reality -- and against the blockchain fundamental design assumption
regarding wallet state and smart contract state --
the off-chain data and information will continue to 
play an important role in asset-related transactions,
and thus the management of such information 
should be considered as a significant component of 
future blockchain-enabled business solutions.

Due to the intrinsic closed design of blockchain systems,
when building business applications beyond crypto-tokens 
(e.g., tokenized currencies), 
it is common to federate actors from a given industry 
to (re)build processes and digital assets models around dedicated purpose-built blockchain systems. 
Such systems run as stand-alone private or consortium networks, 
thus creating several silos of information even within 
the same industry segment and for the same purpose~\cite{IansitiLakhani2017}. 
The challenge of the integration of legacy systems with blockchains 
is an open hot topic and the use of Oracles plays 
a fundamental role in that context~\cite{WEF2020-BridgingGovernance}.

As stated in~\cite{Caldarelli2020-MDPI}, 
the various problems with the current architecture of blockchains emerge whenever there is the need for interactions with the world outside the blockchain system. 
In such cases, trust must be delegated to Oracles, 
which are located outside smart contracts\footnote{We should restate here that the Oracle problem
affects only the cases where smart contracts of a blockchain system 
are required to make decisions based on external sources 
(e.g. stock index, sensors, information managed by external IT systems, 
state of exogenous physical or digital assets etc.). 
Cases like Bitcoin, where the state of the system is 
entirely endogenous, can work without any of the problems addressed here.}. 
Several approaches in implementing Oracles for blockchain-based systems 
have been identified in the literature~\cite{Mammadzada2020}.

Viewing the problem not from a smart contract point of view but from 
a digital asset perspective, 
we can say that the blockchain computational model 
is based on code executed by the platform to manage and maintain 
the state of endogenous digital assets 
(i.e., assets created from within the platform). 
If we revisit the definition of the Oracle problem in that sense, 
we can state the following:
\begin{quote}
Blockchain computation works as a siloed system. 
Smart contracts deployed on a blockchain system 
can only cover computation related to
the management of endogenous digital assets within that (blockchain) system. 
They cannot adequately address business cases involving 
interaction with other systems or management of digital assets 
that can freely flow across systems other than 
that specific blockchain system.
\end{quote}

\section{Our Approach: An Asset-Centric Model}

To solve the apparent conflict between the current smart contract paradigm 
(claiming to be a decentralized and independent compute unit) 
and the fact that smart contracts depend on Oracles, 
we propose a twofold approach:

\begin{itemize}

\item   To consider smart contracts as stateless processing units 
(like stored procedures in traditional Relational Data Base Systems - RDBMS)
that perform the manipulation of the stateful digital twin objects. 
Such smart contracts are executed within a Digital Twin Container (DT-Container). 
During a transaction cycle, the smart contracts request access 
to the digital twins via the DT-Container.
The DT-Container serves copies by-value of the digital twins to the smart contracts.
The smart contracts modify the state of the received digital twins,
and then at the end of processing, they return the modified 
digital twins to the DT-Container for an update. 
The smart contracts cannot duplicate any state of the digital twins nor perform any side-effect computation affecting the state of the digital twins. 

\item   To use the notion of the  ``Design by Contract Specifications''
(DbCS)\footnote{To avoid confusion in 
the terminology where the word ``smart contract'' is used, 
here we use our long phrase ``Design by Contract Specification'' (DbCS)
instead of Meyer's original phrase ``Design by Contract'', 
although the intent is clear.}  
which was developed by Bertrand Meyer 
in the early days of Object-Oriented Programming~\cite{Meyer1991}. 
The basic concept of the DbCS is that the caller of the object-code 
and the provider of the object-code must come to an agreement 
regarding the specification of the inputs, outputs
and the side-effects of invoking the object-code. 
In simple language, there needs to be a binding agreement (aka ``contract'') 
between the caller/invoker of the object-code and 
the object's code
itself (i.e. as defined by its authors).  
In our approach, each time a DT-Container receives an updated digital twin, 
it verifies whether pre/post conditions and invariants are verified and if so, 
it proceeds to persistently modify the digital twin state. 
\end{itemize}

As a result of the above approach, 
smart contracts can be seen as objects solely 
implementing behavioral patterns in the sense of 
the GoF Design Patterns~\cite{DesignPatterns-GOF1994} vocabulary (see~\cite{wiki-gang-of-four} for a short description
of the ``Gang of Four'' (GoF)). 
Structural aspects of the design of a system are 
primarily managed at the level of the DT-Container component, handling the lifecycle of the digital twins. 

We believe that the positive benefits of the smart contracts paradigm 
can be realized without the danger of the siloed data only when 
the focus is correctly placed on the assets. 
That is, we need to view the problem from an asset perspective 
and design the architecture  for smart contracts and blockchains 
to fulfill the requirements of the assets lifecycle. 
These requirements include ease of moveability (migration) of 
assets across new blockchain/DLT systems and legacy systems with the use of digital twins, 
and the privacy of state information associated with digital twins (and their correlated assets).

There are several considerations and assumptions for the design of an Asset-Centric scalable system:
\begin{itemize}[topsep=8pt,itemsep=1pt, partopsep=4pt, parsep=4pt]

\item   {\em Correlation of assets with digital twins}: 
The digital twins are mirroring real-world (digital or physical) assets. 
The digital twins are manipulated either
by smart contracts or traditional software services. 
During their lifecycle, 
the digital twins maintain the strict correlation with the underlying assets.

\item   {\em Persistent state of digital twins}: 
There must be some means 
to persist the state of a digital twin 
that may combine on-chain and off-chain state information. 
Thus, for example, a digitized bearer bond exists both outside the blockchain 
(e.g., in printed sheets with serial numbers in traditional repositories
such as the DTCC~\cite{DTCC-Project-Whitney-2020}), 
while its ownership-state is recorded 
on the blockchain via its correlated digital twin. 
Programmatically, the state is defined by a tuple of (typed) variables.

\item   {\em Consistent syntax for smart contracts expression}: 
Independent of how a smart contract is implemented, 
the syntax for the expression of transition between states of 
the digital twin must be defined in terms of pre/post-conditions and invariants
expressed as type/value constraints on the digital twin variables.

\item   {\em Unambiguous declarative definition of assets}: 
The description of the Digital twin -- in terms of variables and states --
can be done in a declarative way.

\end{itemize}

The main novelty in our approach is that we move the central architectural focus 
from the application logic (implemented either in traditional application services or in smart contracts) to the DT-Container that maintains the digital twins. 
Rather than considering application logic as the core, 
we instead consider that the application logic is unreliable, untrusted, and even potentially written with malicious intent. 

Here, we introduce two important computing constructs that aim 
to maintain the consistency of the assets between the off-chain world 
and the on-chain state digital twin representation.
The first construct is the blockchain DT-Container that interfaces 
between the on-chain and off-chain states of the asset. 
The second construct is the {\em Logical Unit of Work} (LUW), 
which views the processing upon assets to consists of many steps 
(possibly across multiple blockchains) but which must be completed as an atomic unit. 
The consistency of the applications is based on the state of the digital twins --
which can only be updated by their DT-Container if and only if 
their DbCS is verifiable at any moment during their lifecycle. 
The DT-Container guarantees that modifications --
including the transfer of the digital twins across systems 
-- are performed in a controlled way based on a decentralized Logical Unit of Work (LUW) protocol. 
All participants in such a LUW, 
including Applications and DT-Containers are collaborating 
by consistently modifying digital twins using a 
2~phase-commit approach~\cite{Gray1981}. 
The various concepts introduced here are illustrated in the following sections.

\section{Asset-Centric Architectures}
\label{Sec:AssetCentric}

\subsection{Gateways}

The concept of gateways between blockchain systems proposed in~\cite{HardjonoLipton-IEEETEMS-2019}, 
and recently utilized by the Open Data Assets Protocol
(ODAP)~\cite{IETF-draft-hargreaves-ODAP,Hardjono2021-GatewaysBridges}
seeks to address the lack of interoperability between blockchain systems. 
Following the ODAP design, here a {\em Gateway} is a component that 
negotiates transfer of Digital Assets to and from a given blockchain system 
by collaborating with the Gateway of a remote blockchain 
system based on a set of primitives defined by the protocol 
(e.g., the ODAP protocol). 
Gateways are responsible for the complete cycle of 
{\em lock--transfer--commit} 
of digital assets among blockchains.

However, we believe that for the notion of asset transfers 
to reflect a consistent correspondence between the digital asset 
(as represented by ledger-state on the blockchain) 
and the real-world asset as recorded in the off-chain repository 
(e.g., DTCC~\cite{DTCC-Project-Whitney-2020}), 
the design of gateways needs to be extended 
as described in the following sections. 
This extended component implemented in the form of the DT-Container, 
not only covers exchanges among systems (blockchain or otherwise), 
but it also manages consistency between the digital twins, their correlated assets, 
and the processing of their lifecycle by applications 
(smart contracts or traditional software packages).

\begin{figure}[t]
\centering
\includegraphics[width=0.7\textwidth, trim={0.0cm 0.0cm 0.0cm 0.0cm}, clip]{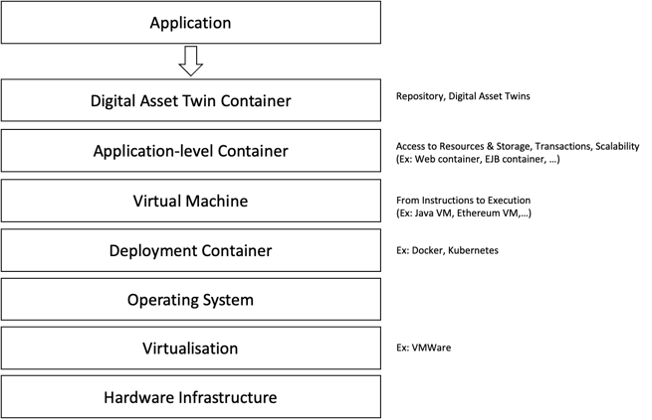}
\caption{Illustration of the various kind of containers emerged over the last decades}
\label{fig:figure2label}
\end{figure}

\subsection{Expanding the Role of Gateways: Digital Twin (DT) Containers}

The notion of a container is not new
and has been used in different contexts in the last few decades. 
In the early 2000s, 
containers were used to refer to application-level coding frameworks,
which typically package technical capabilities such as 
remote invocation, concurrency, load balancing, 
transactionality, authentication, etc. 
Examples of such containers include the Enterprise Java Beans (EJB) Containers 
-- which provides a run-time environment for enterprise beans 
within the application server~\cite{EJB} --
and the Web Container and ``Application Server'' 
containers in general (see Figure~\ref{fig:figure2label}). 
During the 2010s, the containers were also used 
to refer to virtualization capabilities 
at the level of the operating system (e.g., hypervisors),
while more recently the containerization concept has expanded 
to deployment capabilities (e.g., Docker, Kubernetes).

In our case, the notion of Digital Twin Container is instead
a cross-platform software framework that, when deployed on a target system
would offer the same set of technical capabilities regardless whether 
the underlying system is a blockchain/DLT or a traditional application server software stack.

In our definition, 
the Digital Twin Containers must possess the following features or aspects:
\begin{itemize}[topsep=8pt,itemsep=1pt, partopsep=4pt, parsep=4pt]

\item   {\em Clear separation between digital twins and assets}: 
Since assets can be 
in digital or physical form, it is therefore essential to make a clear distinction 
between the asset themselves and their correlated digital twins. 
Digital twins are purely in digital form and are modified by applications 
(smart contracts or traditional software packages). 
Assets -- which may be in non-digital form or 
in digital representation locally --
are managed by specific entities that run specialized software components 
(referred to as {\em Asset Providers} below). 
The legal entities that define and issue assets 
must bear full accountability regarding the various aspects of issuance and management of the assets.

\item   {\em Commitment at the level of Logical Unit of Work}: 
Instead of focusing 
on the unit of transfer as the asset itself (e.g., transferring tokens), 
the commitment protocol must be applied to the LUW in its entirety. 
That is when a gateway accepts the engagement to transfer 
an asset across blockchain networks, 
the gateway is agreeing to provide resources to complete the entire LUW 
that encompasses both the digital twin and the correlated asset. 
Furthermore, the desirable ACID properties 
(atomicity, consistency, isolation, and durability)~\cite{HarderReuther1983} 
are also desirable for current blockchain-based transactions.

\item   {\em DT-Containers as coordinating transaction managers}: 
A further implication of 
the LUW as the unit of commitment between two DT-Containers is the role of the DT-Container 
as transaction manager in the sense of the classic component handling 
distributed transaction processing~\cite{OzsuValduriez2020-ddb-book}. 

\item   {\em Out-of-band applications interactions}: 
The shift of focus to an asset-centric view implies that user interactions occur between applications (i.e., service end-points or APIs) operated by the users. 
The application-to-application interactions occur outside 
and before any actions of the blockchain level.

\end{itemize}

\begin{figure}[t]
\centering
\includegraphics[width=0.9\textwidth, trim={0.0cm 0.0cm 0.0cm 0.0cm}, clip]{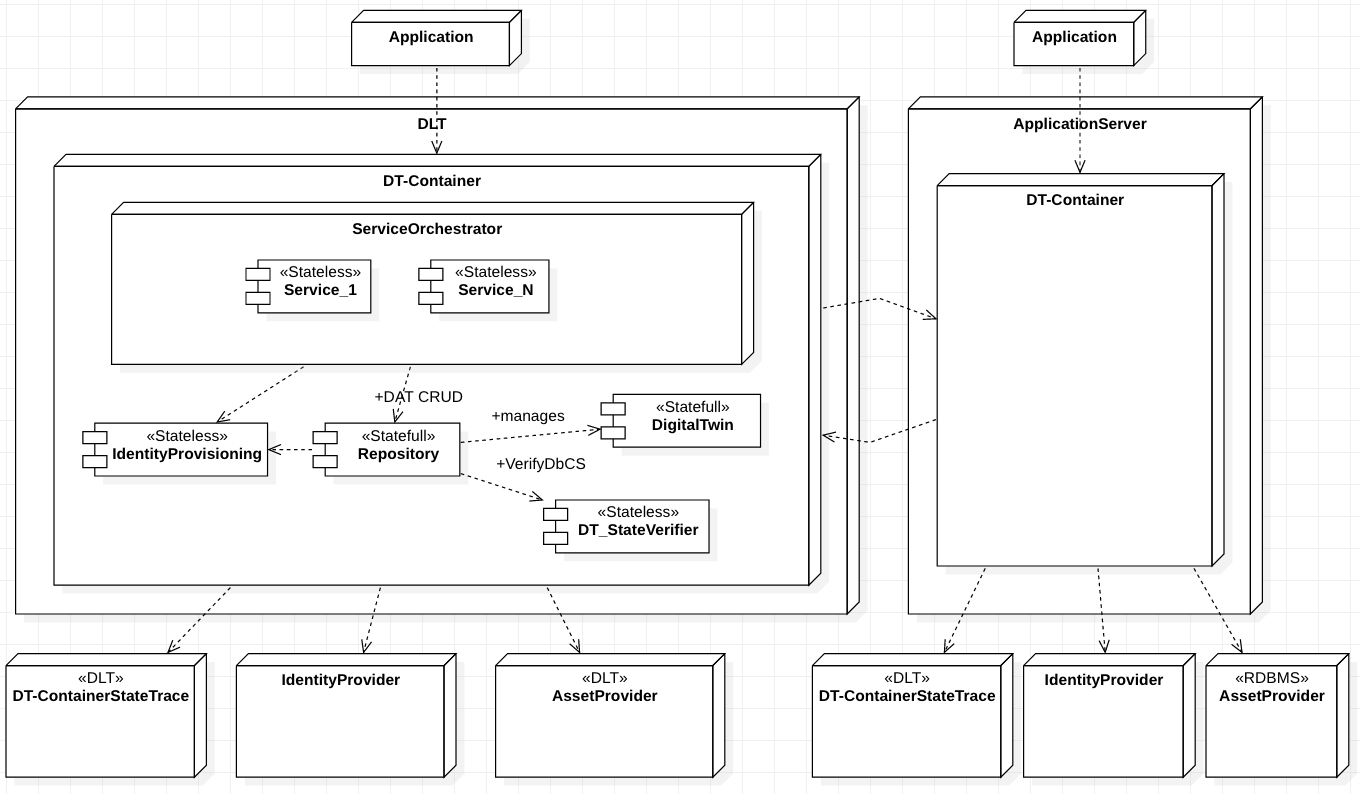}
\caption{Overview of the placement of core business logic away from the smart contracts layer and the validation of outputs of the logic using a DT-Container}
\label{fig:figure3label}
\end{figure}

\subsection{DT-Container components}

As shown in Figure~\ref{fig:figure3label}, 
a DT-Container implements technical requirements 
related to managing the lifecycle of digital twins, 
including the exchange of assets across heterogeneous systems. 
DT-Containers can be deployed directly into a DLT. 
In such case the DT-Container is a set of smart contracts that 
are executed directly by the virtual machine of the DLT. 
In other terms, the DT-Container is a smart contract {\em framework} 
that coordinates execution of other ``business-service level'' smart contracts. 
Saying this, it is also possible to implement a DT-Container 
on a traditional application server, and thus being able to 
manage digital twins on a non-DLT system. 
In such case the DT-Container would be a {\em framework} implemented 
using traditional software (e.g. Java/EJB), 
with assets being managed, for example, in a traditional RDBMS. 
The interaction protocol among DT-Containers would hide the 
internal technological stack and would allow a seamless flow of interaction among DLT and non-DLT systems. 
Irrespective of the technological stack that is used to implement a DT-Container, 
it should always be possible to maintain trace of DT-Container's state 
on a ``witness'' DLT. 
Maintaining a trace of the state of DT-Containers allows for 
internal consistency of the DT-Container state as well as 
for synchronization among DT-Containers in case of failures.

In the specific case of a DLT stack, developers can integrate a DT-Container in the design of enterprise DLT applications in the following manner:

\begin{itemize}[topsep=8pt,itemsep=1pt, partopsep=4pt, parsep=4pt]

\item   They can deploy stateless smart contracts in a smart contract orchestrator. 
The smart orchestrator coordinates the execution of 
the smart contract in accordance 
with the architecture of the underlying virtual machine. 
It offers a framework for the deployed smart contracts to retrieve 
digital twins while confirming any authentication policies related to 
accessing and modifying the state of digital twins (and their correlated assets). 
The smart contract orchestrator is also responsible for 
the consistency of modifications of digital twins in the context of LUWs.

\item   They can implement interaction of applications with {\em Repositories} 
to manage the lifecycle of digital twins. 
The Repository components handle the basic Construct/Read/Update/Destruct (CRUD) 
operations for digital twins. 
The digital twins are defined in a declarative manner 
as state objects with state-transition logic, 
including DbCS governing the transition of states and invariants. A digital twin {\em verifier} of the DbCS is responsible for the
global consistency for every state transition of the digital twin 
requested by a smart contract.

\item   They can integrate with identity providers via the {\em Identity Provisioning} module
to obtain authentication credentials: The Identity Provisioning module 
implements standardized interfaces with Identity Providers that are able 
to authenticate users (e.g., via standard authentication and authorization protocols,
such as Kerberos~\cite{SteinerNeuman1988}, 
OpenID-Connect~\cite{OIDC1.0}, 
{SAML2.0-SSO}~\cite{SAMLwebsso} 
and newer schemes such as DID methods~\cite{W3C-DID-2018}).
This permits the verification of whether modifications of digital twin states
have occurred  according to policies declared in the definition of 
the state transition of digital twins
(e.g., must be performed only by authenticated and authorized entities).

\item   They can benefit from {\em Asset Providers} 
that are connected to the DT-Container via Repositories and which
manage real-world assets (physical or digitized): Asset Providers are software components 
that act as ``custodian'' of assets. 
The legal entity that runs an Asset Provider component bears 
the legal responsibility for issuing and managing assets. 
In terms of technical constraints Asset Providers are required to 
implement ACID transactions for assets (eventually by escrowing assets for the duration of LUWs). 
That way, they can ensure the correct correlation of their assets with 
the corresponding digital twins that participate in LUWs.

\end{itemize}

\section{DT-Container Mediated Access}

One of the key roles for DT-Containers is to mediate between 
independent systems/networks and to hide 
the complexity of the internal systems from outside entities. 
In the context of our asset-centric model for smart contracts and blockchains, 
we believe that DT-Containers provide a suitable means to achieve 
the desirable goals stated in the previous section. When a DT-Container is deployed in a blockchain system, it ensures that a consistent state is achieved before and after the target smart contract on the blockchain is invoked. 
That is, the DT-Container must ensure that there is agreement and synchronization 
between the behaviors of on-chain smart contracts and the inputs, outputs,
and side-effects of the underlying assets. 

More specifically, DT-Containers must satisfy the following: 
\begin{enumerate}[topsep=8pt,itemsep=1pt, partopsep=4pt, parsep=4pt]

\item   Must provide full synchronization of digital twins and real-world assets: 
In order to do so, it must be possible for applications to verify that assets are reserved and can be committed upon successful completion of an LUW. 
Digital twins should be fully correlated with their assets during their whole lifecycle. 
Direct verification (for example through API calls to asset providers) 
should be possible at any time by applications, 
and it can be decisive to commit or roll-back an LUW.

\item   Must allow asset verification for local as well as for remote systems: Although the blockchain can guarantee the persistence of state data 
on the ledger replicated by all its nodes, 
applications may not have access to the state of digital twins and underlying assets. DT-Containers provide mediation of access to asset state whether such assets are managed in the local system or in a remote system. 
In the latter case,
communication among DT-Containers would facilitate access 
to the state of digital twins and assets participating in the same LUW.

\item    Must provide the interface for off-chain systems to invoke on-chain smart contracts: 
Another mediating role of a DT-Container is to provide 
an interface for off-chains (non-blockchain) systems 
to invoke specific types of smart contracts indirectly. 
This may be an important function for use-cases of private/permissioned blockchains. In these cases, the DT-Container validates authorship and code-safety and can invoke the smart contract on behalf of external systems.

\end{enumerate}

In summary, 
DT-Containers coordinate their actions to implement decentralized 
transactions in each participating system (blockchain or legacy). 
Successful completion of the LUW is achieved when all individual transactions in each system are completed in a successful manner.
Otherwise, all operations are rolled-back, 
and all digital twins participating in the LUW are reverted 
to their original state at the beginning of the LUW.

\section{Architectural implications}

\subsection{Implications for smart contract virtual machine design}

In the approach adopted by programmable blockchain systems so far, 
smart contracts are the central component of the design.
The Oracles act as auxiliary components supplying smart contracts 
with information from the external world. 
In our approach, the Digital Twins and related {\em Repositories} 
play the main role by guaranteeing that Digital Twins and their correlated Digital Assets managed by {\em AssetProviders} are consistent at any point in time. 

In our case the smart contracts are processing units 
that obtain Digital Twins from Repositories, perform actions with the intent of updating the state of these Digital Twins, 
then request for update of the state. Repositories are strictly managing updates of the state of Digital Twins. 

Smart contracts are effectively stateless process units. In the case the DT-Container is deployed on a DLT system, the DT-Container and the {\em SmartContractOrchestrator} 
(see Figure \ref{fig:figure3label}) are themselves smart contracts 
that act as frameworks that implement patterns to coordinate interactions 
among smart contracts deployed by developers. 
Much like Enterprise Java Beans (EJBs) deployed in a EJB container, 
smart contracts deployed by developers run inside the {\em SmartContractOrchestrator}. 
All interactions with Repositories, Digital Twins, 
as well as exchanges with applications from remote systems go through the DT-Container layer.

\subsection{Implication for cloud service development}

The notion of application server container is quite common 
in traditional web applications. 
Beyond traditional services like transactions management systems 
or user identification management, 
DT-Containers for non-blockchain SaaS systems (Software-as-a-Service)
allow for the implementation of a standard mechanism 
to communicate with DT-Containers 
in remote blockchain systems and vice-versa. 
The inter-DT-Container interactions must be standardized 
in order to permit Logical Units of Work (LUWs) to spawn across these systems
(i.e., across traditional legacy systems and blockchain-bases systems).

\subsection{Implication for Asset Providers}

Asset providers must implement ACID behavior in the management of their Assets. 
In that sense, the asset providers are resource managers 
with the responsibility to make assets available 
to applications via DT-Containers. 
Asset Providers can range from databases or 
Enterprise resource planning (ERP) systems,
to smart contracts delivering on-chain digital assets 
(e.g., Ethereum ERC-20 or ERC-721 smart contracts). It is the responsibility of the Asset Provider to comply with the interaction pattern defined by the DT-Container to maintain the correlation between Digital Twins and Assets during computations in the context of logical units of work (LUWs). {\em Repositories} are interacting with Asset Providers to commit or roll-back resources based on the status of the LUW.

\subsection{Centralization vs decentralization}

Decentralization introduces absence of central control in the technical architecture of systems. 
However, a clear distinction must be made between accountability and liability (which are centralized by essence) 
and the decentralized software architectures. 

For use cases that involve legal entities or public institutions 
it is unlikely that responsibility and accountability
will be delegated to programmers of smart contracts or software applications 
without strict control in terms of the legal implications related to offering assets to users. 
The issuance of assets is by definition centralized.
Even in the case of the Decentralized Autonomous Organization (DAO)
project contention~\cite{Siegel2016} (the resolution of which resulted
in the fork within the Ethereum blockchain~\cite{Castillo2016}),
the decision to carry-out the fork was in reality
made by a small group of people in a centralized manner.

The DT-Container is not thus a component that defies 
the decentralization principle of blockchain. 
Instead, it makes explicit the basic fact that digital assets 
must be managed under clear responsibility by well-identified parties.

\section{A note on Asset and Owner Identification}

With the rise of blockchain technology, 
cryptocurrencies and the use of public-keys as a means to transact on the blockchain,
there has been considerable confusion in the industry regarding the 
matter of digital identity~\cite{Hardjono2019-IEEECommsMagazine-short}.  
Among others, some confusion exists relating to 
(a) the authority to designate (assign or allocate) namespace ownership 
and the identifiers within the namespace, 
(b) the day-to-day control over the namespace (and any identifiers within it), and 
(c) the authority to bind a public-key to an identifier under a given namespace.

The source of authority in binding a public-key to an identifier 
(e.g., subject's name) is crucial in digital transactions 
using the public-key (both on-chain and off-chain).
Parties in a bilateral transaction typically wish 
to obtain assurance that they are dealing with the correct entity 
(possessing the correct public-key) in the transaction. 
They wish to reduce risk, 
and therefore require that risk to be allocated to (assigned to) 
the entity that performs the binding of the public-key to the identifier. 
This role as the binding authority has traditionally been 
performed by the Certification Authority (CA) entities, 
who typically operate under a contractual service level agreement (SLA) 
referred to in the industry as the Certificate Practices Statement (CPS)~\cite{RFC2459-formatted}.  
The rudimentary CPS is defined in~\cite{RFC3647-formatted} for 
{X.509} certificates.

In the context of digital twins, each digital asset (non-human asset) must be assigned a globally unique identifier string (under a unique namespace) by a naming authority. 
For example, for certain categories of financial assets, 
an International Securities Identification Number (ISIN) 
consisting of a 12-digit alphanumeric code must be assigned to the asset. 
The agency within each country with authority over its ISIN namespace is the country's respective National Numbering Agency (NNA). 
This single source of naming authority has the benefit 
that there is perpetual consistency between: 
(i) the digital asset in the cyberworld (recognized via its digital identifier, such as the ISIN number),
and (ii) the real-world asset that may be physically located 
in a certain part of the world (e.g. paper certificates of ownership in repositories, 
such as in the DTCC in New York).

Recently, with the emergence of blockchains and DLT systems,
the notion of a ``self-sovereign'' identity~\cite{IEEE-Comms-Identity-SSI} 
has come to the forefront as a means for individuals to obtain control 
over their digital identities. 
This desire is not new, and it is as old as the Internet itself.
The core concept in SSI is that a permissionless blockchain system 
would permit an individual to declare ownership of a given 
public-key pair by way of ``publishing'' -- to the decentralized ledger of the blockchain --
a binding between the public-key under the control of the individual and the name-string of the individual.  
Here, publishing is taken to mean that the matching private-key
would be used to sign the blockchain transaction, 
thereby providing to the community a proof-of-possession (PoP) of the private-key. 
Since the ledger is replicated at all nodes, 
unauthorized modifications to the binding would be infeasible. 
Furthermore, ``self-sovereignty'' is thought to have been 
achieved because the individual becomes the sole entity 
who can (at any time) update the ledger of the blockchain with newer bindings.
Data structures to represent this binding have been developed~\cite{W3C-DID-2018}.

We believe the SSI approach -- like the PGP approach~\cite{RFC1991-formatted} 
of the 1990s --
confuses the basic notions of (a) control over a public-key pair
and (b) the accountability and responsibility of utilizing the key pair.
The decentralized ledger may provide an efficient mechanism
to achieve (a), but in business transactions the existence
decentralized ledger does not absolve the individual from (b).
In most online business transactions involving public keys
the counter-party wishes to obtain assurance
that the binding between a public-key and the subject legal name
is correct and truthful.
A self-asserted public key -- as in the case of the PGP or SSI schemes --
will be unacceptable to parties in a transaction
because of an unfair burden of risk
(e.g., one or both parties lie about their bindings, 
or one or both claim loss of their private-key immediately 
after signing a business agreement and repudiating the agreement).
It is for this reason that the PGP model was never 
adopted by the business community.

It is also for this reason that the Certification Authority (CA) 
segment of the computer industry
grew considerably starting in the mid-1990s - despite
criticisms about the CA entities being ``centralized''.
There was a business demand for a trusted third party
to take on the liabilities of managing the X.509-based
bindings between a public-key and its legal owner 
(i.e. the subject stated in the X.509 certificate).
A major role of the CA entity is to 
undertake legal liability for the binding and 
to provide warranties in the case
where an identity/key binding proved to be false or erroneous.
The CA provided the mechanism for users to report key losses quickly,
so that a stolen public key can immediately 
be placed on the certificate revocation list (CRL).
The CRL database is accessible globally via standardized protocols and APIs 
(i.e., the Online Certificate Status Protocol and 
OCSP-Responders~\cite{RFC6960-formatted}).
The role of the CA was thus to reduce 
the risk on the part of the transacting parties.

\begin{figure}[t]
\centering
\includegraphics[width=1.0\textwidth, trim={0.0cm 0.0cm 0.0cm 0.0cm}, clip]{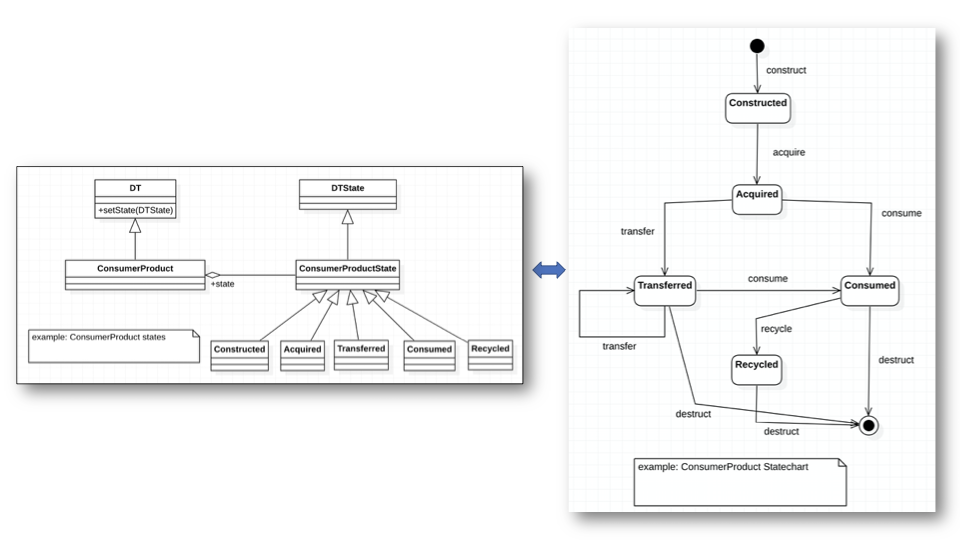}
\caption{Digital Twin: Digital Twin state and stateless smart contract specification}
\label{fig:figure4label}
\end{figure}

In the context of the architecture proposed in this paper,
we explicitly and deliberately make a clear distinction between 
the globally unique identifier of a digital twin and 
any digital identity mechanism used to refer 
to ownership and accessibility rights on a digital twin. 
The DT-Container architecture provides a bridge to connect a digital twin 
with both an identity provider and an asset provider 
as two interconnected but clearly distinct parties.

\section{Example}

To illustrate the concepts introduced in this paper we use a real-life implementation using the {\em Compellio Registry} middleware\footnote{https://compell.io/products/blockchain-registry}. 
The Compellio Registry middleware acts as a bridge between applications 
and systems managing assets (digital or physical). 
It implements the DT-Container features, 
acting as a middleware between legacy and blockchain systems 
where assets managed by the underlying systems are correlated to Digital Twins.
Applications interact with the middleware to fetch Digital Twins and to consistently modify their states. 
Digital Twins participate in LUWs and the middleware takes care of related cross-system transactional coordination by modifying the state of Digital Twins across systems in a consistent manner. 

The example presented below is borrowed from the domain of consumer product traceability and authentication. In the specific example,
the digital twins are correlated to physical products.

\subsection{Digital Twin state}

As shown in Figure~\ref{fig:figure4label}
the digital twins are defined in a declarative way. 
A {\em ConsumerProduct} is a digital twin that follows 
a life-cycle defined by various states ({\em ConsumerProductState}).

\begin{figure}[t]
\centering
\includegraphics[width=0.9\textwidth, trim={0.0cm 0.0cm 0.0cm 0.0cm}, clip]{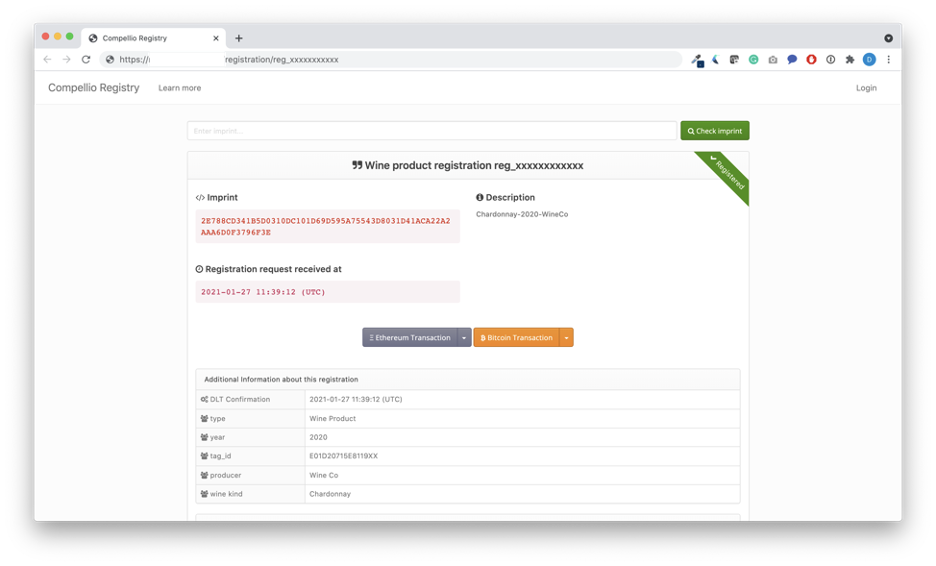}
\caption{Digital Twin: view and verification (human or machine) as constructed state}
\label{fig:figure5label}
\end{figure}

Figure~\ref{fig:figure5label} shows a visual representation of 
a digital twin on Compellio Registry middleware. 
In this specific case, the digital twin represents a ConsumerProduct instance 
(and more specifically, a wine bottle in the example). 
The state of the digital twin is defined by a set of attributes. 
The attributes are in the form of \texttt{<key, type, value, scope>} 
where scope is defined by three possible values:
(i) \texttt{private}, i.e., attributes that are accessible by the owner of the digital twin; 
(ii) \texttt{restricted}, i.e., attributes that are accessible by users explicitly authorized by the owner of the digital twin; 
(iii) \texttt{public}, i.e., attributes that are accessible by any user. 
In the example below the attributes of the digital twin 
are related to the product 
(type of product, producer, year, wine kind, IoT ``tag ID'', etc.). 
Figure~\ref{fig:figure5label} shows only public attributes. 
Users with special permissions can access protected and private attributes. The exact structure of the attributes and states of a digital twin can be formally defined by specific ontologies or other schema-driven definitions.  

Note: As a special case, attributes can refer to other digital twins. 
In such cases, the attributes represent the relations 
to digital twins (referred by dt\_ref type) already created in the system. 
See for example, attributes like acquired\_by or consumed\_by that 
are relationships to the digital twins referring to users.

\subsection{Digital twin state registration}

\begin{figure}[t]
\centering
\includegraphics[width=0.7\textwidth, trim={0.0cm 0.0cm 0.0cm 0.0cm}, clip]{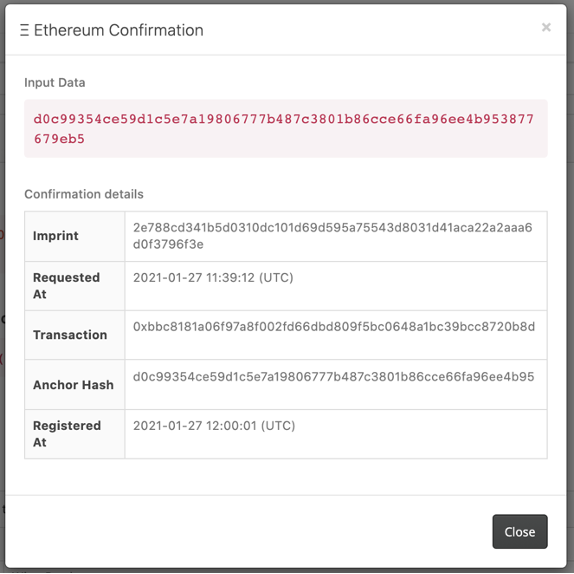}
\caption{Digital Twin: view and verification (human or machine) as constructed state}
\label{fig:figure6label}
\end{figure}

When a digital twin enters a new state, the state is registered on-chain. 
Figure ~\ref{fig:figure6label} illustrates the details of the 
on-chain confirmation of the digital twin state registration.

\begin{figure}[t]
\centering
\includegraphics[width=0.9\textwidth, trim={0.0cm 0.0cm 0.0cm 0.0cm}, clip]{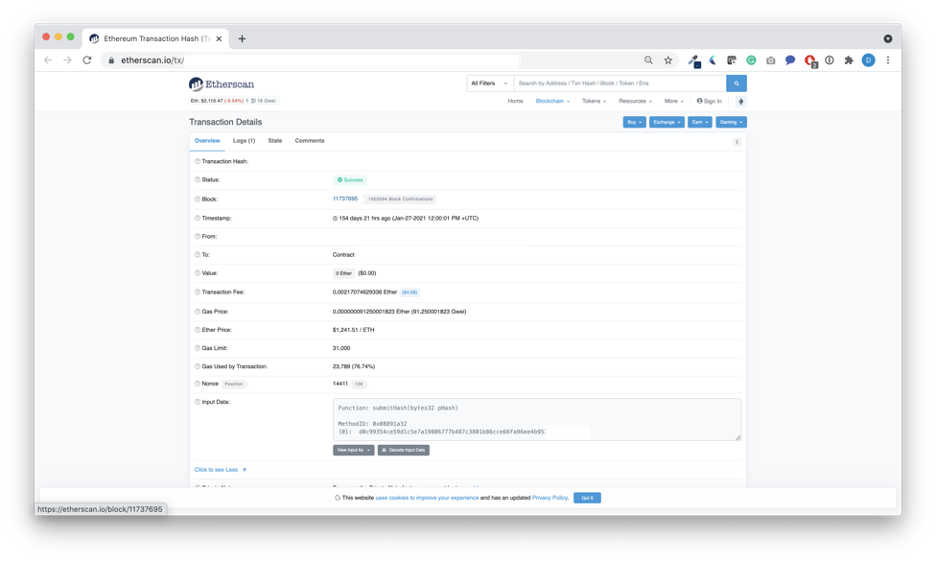}
\caption{Digital Twin: Ethereum registration transaction}
\label{fig:figure7label}
\end{figure}

Figure~\ref{fig:figure7label} illustrates the Ethereum transaction 
related to the registration of the {\em constructed} state of the digital twin. 
The digital twin gets serialized and the hash in registered on-chain 
via that transaction. 
The hash anchored on-chain appears in the {\em Input Data} attributes of the transaction. 

Note: for confidentiality reasons,
the hash of the serialized differs from the hash anchored on chain. 
Several ways of maintaining the link between the state hash and 
the anchored hash may be implemented, including for example, a timestamp~\cite{HaberStornetta1991,BayerHaber1993}.

\subsection{Digital Twin - state transition}

When an application which is deployed on the DT-Container 
(e.g., smart contract managed by the Smart Contract Orchestrator or a software component deployed in the DT-Container of a non-blockchain Application Server) 
seeks to update the state of a digital twin, 
it calls the Repository component of the DT-Container by passing a new state 
to the digital twin. 

The Repository component then fetches the current state of the specific digital twin instance (\texttt{dat\_ref\_12345678} in the example below) and goes through verification of the \texttt{DbCS} state transition via the DT\_StateVerifier. 

\begin{figure}[t]
\centering
\includegraphics[width=0.9\textwidth, trim={0.0cm 0.0cm 0.0cm 0.0cm}, clip]{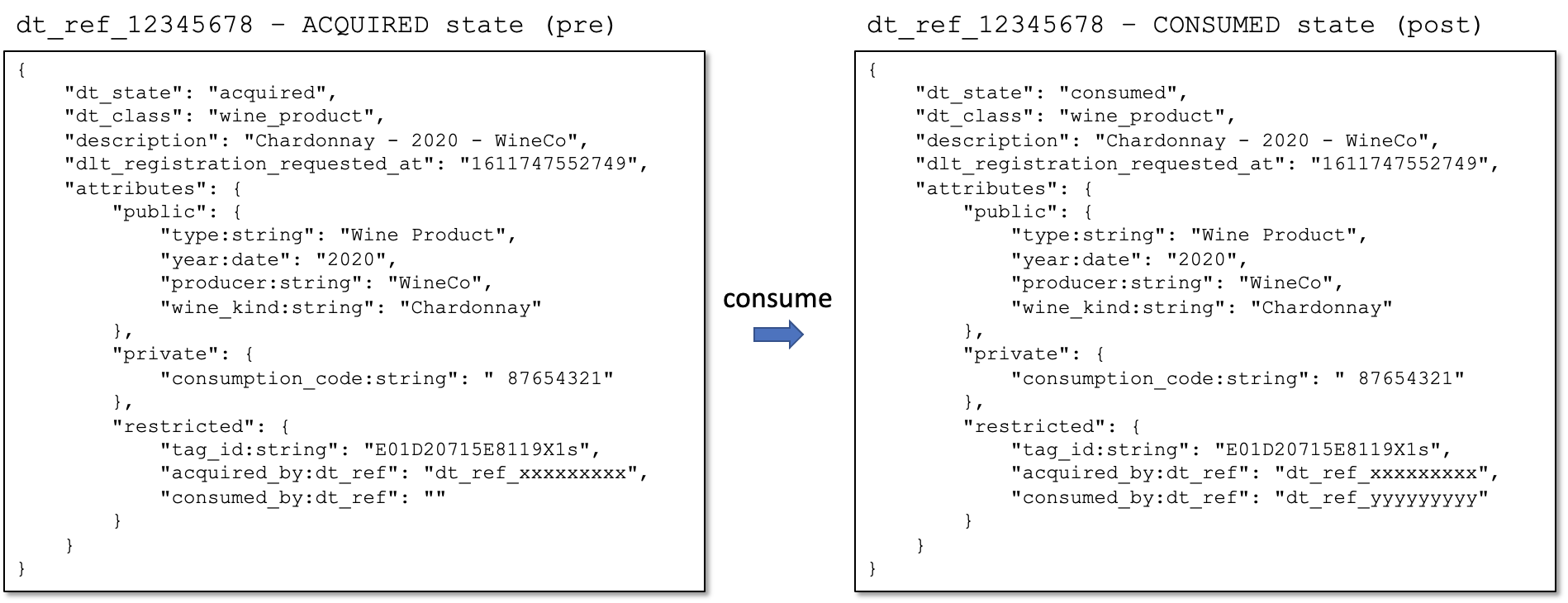}
\caption{Digital Twin transition between states}
\label{fig:figure8label}
\end{figure}

Lets suppose that an application requests transition of 
the digital twin from state ``Acquired'' to state ``Consumed''. 
The application requests to the Repository for the transition 
to the Consumed state, by way of passing a serialized state. 
Figure~\ref{fig:figure8label} illustrates such a transition. 
For the transition to be valid,
the user that consumes the product must supply 
a consumption-code that is equal to the value of 
the consumption code attribute of the digital twin 
(nb. here we omit the UX process/details on how the operation is performed).

\subsection{Logical unit of work}

\begin{figure}[t]
\centering
\includegraphics[width=0.8\textwidth, trim={0.0cm 0.0cm 0.0cm 0.0cm}, clip]{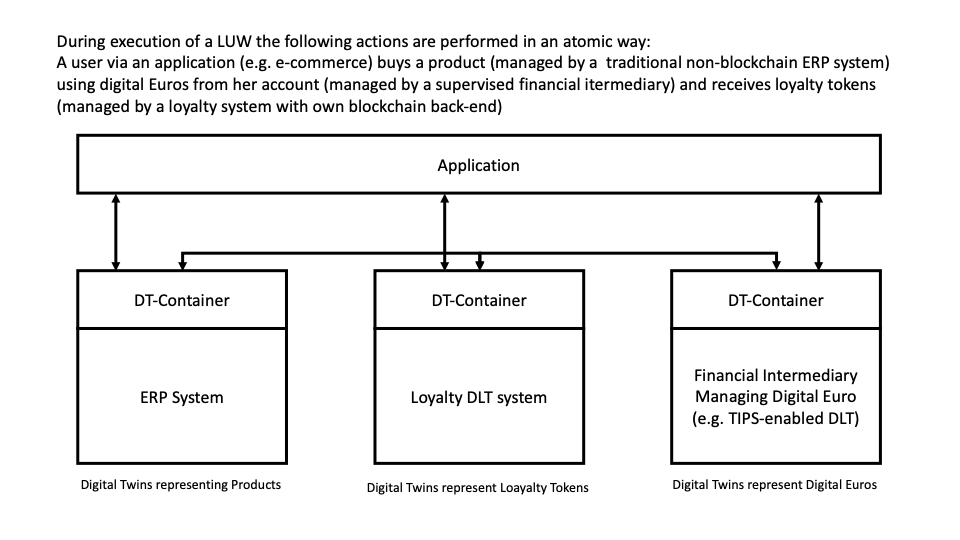}
\caption{Example of a LUW managed by DT-Containers involving coordination among both legacy and blockchain systems}
\label{fig:figure9label}
\end{figure}

A Logical Unit of work manages ACID transactions among systems, where the Digital Twins and their underlying Digital Assets must be modified following an all-or-nothing principle.
Figure~\ref{fig:figure9label} illustrates a case where DT-Containers 
are used to implement simultaneous execution of actions during acquisition 
of product, 
i.e., transition of the digital twin representing a product to the \text{acquired} state. 

In this example we assume following: 

\begin{enumerate}[topsep=8pt,itemsep=1pt, partopsep=4pt, parsep=4pt]

\item The producer manages its inventory and integrates e-commerce transactions using traditional ERP software. 

\item The producer is also using a customer loyalty service,
offering digital tokens via a DLT.

\item   The user holds an account in Digital Euros 
on a supervised financial intermediary 
(e.g., following a hybrid bearer digital euro and 
account-based infrastructure presented in~\cite{ECB2020-DigitalEuro}). 

\end{enumerate}

In the proposed architecture as shown in Figure~\ref{fig:figure9label}, 
the DT-Containers are integrated with each of the three systems above. 
For the purpose of an example,
assume that an e-commerce application implements 
a LUW -- in which case the following actions 
must be performed in an atomic way:
\begin{itemize}[topsep=8pt,itemsep=1pt, partopsep=4pt, parsep=4pt]

\item The product must be acquired 
(i.e., the digital twin moves to state \texttt{acquired} and the user is notified of its acquisition);

\item The account of the producer is credited with digital euros for the value of the product;

\item The loyalty account of the user is credited by the producer with 
some loyalty token, in accord with the customer loyalty program in place . 

\end{itemize}

To perform this operation,
the application initiates the LUW and makes use of 
the digital twins of the various systems. 
When all the above actions are performed 
the application requires completion of the LUW. 
At that moment, the DT-Containers are coordinating the consistent update of the underlying assets while ensuring that the correlated digital twins reflect the exact state of the assets.

\section{Related Work}

In the past few years, there has been considerable interest and attention
placed on the challenge of {\em atomic swaps}
between blockchain systems,
focusing primarily on permissionless (public) blockchains.
Many atomic swaps schemes view the problem of digital asset management
from a blockchain-centric perspective,
ignoring the fact that in many use-cases the real-world assets
truly exists outside the blockchain domain
and that the smart contracts paradigm as defined today
fall short of being the coordinating point between
the real-world and the digital world.

The basic notion in atomic swaps
is the ``swap'' (trade or exchange) of one virtual asset to another
in a concurrent interconnected manner.
The basic two-party {\em atomic swap} using hash-locks in permissionless blockchains
appears to have been first proposed in~\cite{Nolan2013}.
There are a number of issues around the notion of atomic swaps
as defined using hash-locks in~\cite{Nolan2013,BIP199-formatted,BlackLiu2018-ERC1630}.
Firstly,
each permissionless blockchain may have 
different throughput rates 
(i.e., block confirmation rates/speeds) based on the consensus protocol
employed by each of the blockchains.
This means that even with hash-locks and time-locks,
additional synchronization mechanisms must be employed between blockchains
(e.g., a centralized certified blockchain in~\cite{HerlihyLiskov2019};
witness blockchain in~\cite{ZakharyAgrawal2019}).
Secondly,
many atomic swap schemes assume that the ``value'' of the assets does not change dramatically during the swap and settlement periods in the respective blockchains. However, the difficulty lies in how to guarantee this value stability. Proposals such as~\cite{ZamyatinHarz2019-XCLAIM} are effectively forced 
to introduce adjunct functions/services to address the value-stability problem.

The limitations of the smart contracts paradigm have also been discussed in~\cite{Herlihy2019-CACM}. One of the core ideas underlying the notion of smart contracts is the sharing of function and state information,
where the ledger provides a universal sharing and accessibility to the shared state information. 
Both of these computing constructs are not new inventions ushered in by the blockchain revolution, but rather originate 
from the field of object-oriented programming 
(i.e., methods and private/public variables)~\cite{DickersonHerlihy-PODC2017,Zakhary2019Global}.

Related to our proposed use of DT-Containers for coordination, 
the Chainlink design offers a set of on-chain and off-chain 
components that allow smart contracts to interact with external sources. 
Chainlink provides smart contracts with the ability to push and pull data, facilitating the interoperability between on-chain and off-chain applications~\cite{Harper2018-Chainlink}. Off-chain nodes are responsible for collecting the data from the off-chain resource as requested by user contracts. After retrieving the relevant data, these nodes process that data through ChainLink Core, namely the core node software that allows off-chain infrastructure to interact with ChainLink's blockchain. Once the data is processed, ChainLink Core transmits it to the on-chain oracle contract for result aggregation~\cite{Harper2018-Chainlink}. In our approach, we call for a fundamental modification of on-chain computations, introducing the notion of logical units of work and design by contract specifications as two foundation concepts in smart contract programming.

With regards to the need for the separation between structure from computation, the work on Obsidian~\cite{Coblenz2017-Obsidian}
follows a similar approach to our current work, 
by making resources hold by a smart contract an explicit construct. 
Obsidian models the resources with linear types, 
which statically restrict the life-cycles, 
so that resources cannot be accidentally lost. 
Obsidian is a typestate-oriented language, 
representing state in types and statically preventing some invalid invocations.
In our approach, the declarative definition of Digital Twin states 
follows the same objective by ensuring via post-conditions 
that computations maintain Digital Twins in consistent states.

\section{Conclusions}

In the current work we have proposed a more asset-centric view of blockchain systems and distributed ledger technologies, placing these technological components in their proper place within the broader computerized digital assets ecosystem.

Our novel contribution is that we move the central architectural focus from the application logic -- implemented either in traditional application services or in smart contracts -- to the Digital-Twin Container, which is one of the computing constructs aimed at maintaining the consistency of the assets between the off-chain world and the on-chain state digital twin representation. 
The notion of the Digital Twin Container is a cross-platform software framework 
that when deployed on a target system
would offer the same set of technical capabilities regardless whether 
the underlying system is a blockchain/DLT or a traditional application server software stack.

The construct of the Logical Unit of Work (LUW) is used as 
the unit of processing of asset-related transactions and which must be completed atomically. 
The successful completion of a given LUW 
is achieved when all individual transactions in each system 
have been completed in a successful manner.
In this sense, we rediscover and borrow from the proven paradigm of containers that originated from over two decades ago in the area of enterprise applications and which today have expanded into the containerization concept deployed currently in most cloud deployment infrastructures.

We expanded the role of cross-blockchain gateways to incorporate the DT-Container, 
and described an architecture that shifts core technical coordination capabilities into the portable DT-Containers. 
We thus simplify smart contract design into a layer that focuses on business logic applied to digital twins. 
DT-Containers -- in collaboration with asset providers -- handle all aspects related to consistency among digital twins and their correlated (physical or digital) assets.

~~\\




\end{document}